\documentclass[letterpaper,aps,showpacs,floatfix,11pt,prc]{revtex4-1}
\usepackage{graphicx}
\usepackage{amsmath,amssymb,amsbsy,bm}
\usepackage{graphicx}
\usepackage{comment}
\usepackage{float}
\usepackage[colorlinks=true,linkcolor=blue,citecolor=blue,urlcolor=blue]{hyperref}

\begin{document}

\title{Evolving Charge Correlations in a Hybrid Model with both Hydrodynamics and Hadronic Boltzmann Descriptions}
\author{Scott Pratt}
\affiliation{Department of Physics and Astronomy and National Superconducting Cyclotron Laboratory\\
Michigan State University, East Lansing, MI 48824~~USA}
\author{Christopher Plumberg}
\affiliation{Theoretical Particle Physics, Department of Astronmy and Theoretical Physics\\
Lund University, S\"{o}lvegatan 14A, SE-223 62, Lund, Sweden}.
\date{\today}


\begin{abstract}
Correlations related to local charge conservation provide insight into the creation and evolution of up, down and strange charges in the quark-gluon plasma. Here, the evolution of charge correlations is overlaid onto a hydrodynamic calculation for the regions where temperature exceed 155 MeV, then transferred and carried through a microscopic model of the hadronic stage. Thus, for the first time, charge correlations are evolved consistently with a full state-of-the-art description of a heavy-ion collision. The charge correlations are projected onto charge balance functions, which characterize such correlations in the final state, and are presented as a function of relative rapidity and relative azimuthal angle for Au/Au collisions. The role of the hadronic stage is investigated. Calculation of the contribution to charge-separation observables related to the chiral magnetic effect are also presented. Calculations are compared to data from the STAR Collaboration at RHIC (Relativistic Heavy Ion Collider) data when possible. 
\end{abstract}

\maketitle

\section{Introduction}
\label{sec:intro}
Relativistic heavy ion collisions are characterized by copious charge production. During the initial stage, assuming that a chemically equilibrated quark-gluon plasma (QGP) is created, roughly equal numbers of up, down and strange quarks are produced. Within the central unit of rapidity, more than a thousand hadrons might be emitted in a single collision, with each hadron carrying two or three quarks. Thus, during a single central collision, a rich assortment of up, down and strange charge is created, organized into hadrons and emitted. Given that the average charge density is zero, because there are nearly as many antiquarks as quarks, one cannot well characterize chemical properties by the densities of conserved charges. Instead, the chemical properties of the medium are best reflected by the susceptibility, which is a measure of the charge fluctuation. For an equilibrated system away from the conditions of phase separation, the charge correlation $C_{ab}(\vec{r}_1,\vec{r}_2)$ should be local, with its strength determined by the susceptibility $\chi_{ab}$,
\begin{eqnarray}
\label{eq:Cchi}
C_{ab}(\vec{r}_1,\vec{r}_2)&=&\langle \Delta\rho_a(\vec{r}_1)\Delta\rho_b(\vec{r}_2)\rangle\\
\nonumber
&=&\chi_{ab}(\vec{r}_1)\delta(\vec{r}_1-\vec{r}_2)\\
\chi_{ab}&=&\frac{1}{V}\langle \Delta Q_a\Delta Q_b\rangle,\\
\nonumber
\Delta\rho_a(\vec{r})&=&\rho_a(\vec{r})-\langle\rho_a(\vec{r})\rangle,\\
\nonumber
\Delta Q_a&=&Q_a-\langle Q_a\rangle.
\end{eqnarray}
Here, the indices $a$ and $b$ refer to the charge species, up, down and strange.  For the purposes of this study, the delta function is any short-range function that integrates to unity, unless one needs to view the correlation on extremely small length scales, $\lesssim 1$ fm. In a weakly interacting QGP, the range of the correlation is effectively zero, while in a hadron gas the correlation extends over the size of hadron. Near a phase transition, the correlation length can grow arbitrarily, but for systems with small net baryon number, lattice shows no evidence of a phase transition, and correlations lengths are expected to be small \cite{Borsanyi:2011sw}. For the highest collision energies at the Relativistic Heavy Ion Collider (RHIC) or in collisions at the Large Hadron Collider (LHC), the net charge density approaches zero, and the quantities $\Delta\rho$ and $\Delta Q$ can be replaced by $\rho_a$ and $Q_a$, a simplification applied throughout the paper.

In the gaseous Quark-Gluon Plasma (QGP) limit, quarks are independent of one another, and $\chi_{ab}$ is diagonal,
\begin{eqnarray}
\chi_{ab}^{\rm QGP}&\approx& (n_a+n_{\bar{a}})\delta_{ab},
\end{eqnarray}
where $n_a$ is the density of quarks of type $a$. In a hadronic gas, the correlation can be diagonal as a hadron species $h$ can have multiple charges,
\begin{eqnarray}
\label{eq:chih}
\chi_{ab}^{\rm had}&\approx&\sum_h n_qq_{ha}q_{hb},
\end{eqnarray}
where $q_{ha}$ denotes the charge of type $a$ on a hadron of species $h$. In between these two limits, $\chi$ is complicated, and has been calculated by lattice gauge theory, which shows a smooth transition between the expression for a hadronic gas to one for a QGP in the temperature range $150<T<225$ MeV \cite{Borsanyi:2011sw,Bellwied:2015lba}. Thus, validating that one has indeed created matter with equilibrated chemical properties requires verifying that the susceptibility, or the local part of the correlation, varies with time and position according to the local temperature.

Because of local charge conservation, any local correlation such as one would expect in the hadronic breakup stage characterized by a delta function, as in Eq. (\ref{eq:Cchi}), and with the susceptibility such as that for a hadron gas in eq. (\ref{eq:chih}),  must be accompanied by a balancing correlation.  In the context of a relativistic heavy-ion collision, charge conservation requires that the net charge correlation, summing both the short-range and the longer-range balancing contributions, integrates to zero.  If one defines the balancing contribution to the correlation as $C'(\vec{r}_1,\vec{r}_2,t)$,
\begin{eqnarray}
\label{eq:Cprimedef}
C_{ab}(\vec{r}_1,\vec{r}_2,t)&=&\chi_{ab}(\vec{r}_1,t)\delta(\vec{r}_1-\vec{r}_2)+C_{ab}'(\vec{r}_1,\vec{r}_2,t).
\end{eqnarray}
If the system is locally equilibrated, $\chi_{ab}$ in Eq. (\ref{eq:Cprimedef}) is indeed the equilibrated susceptibility as calculated in lattice gauge theory. Due to charge conservation $C_{ab}$ must integrate to zero, and
\begin{eqnarray}
\label{eq:ccons}
\int d^3r' C'_{ab}(\vec{r},\vec{r}',t)&=&-\chi_{ab}(\vec{r},t).
\end{eqnarray}
In a hadronic state, the non-local charge correlation represents the inter-hadron charge correlation. As $\chi_{ab}$ changes with time, it must feed $C'(\vec{r},\vec{r}',t)$ at $\vec{r}=\vec{r}'$ due to the local nature of charge conservation. Given that the local correlation, the correlation between charges on the same hadron, are determined by the various hadronic yields, it is the determination of the balancing correlation $C'_{ab}$ that carries new information. If charges are created early, perhaps in the pre-hydrodynamic stage of the reaction, the correlation $C'_{ab}$ has the chance to spread over a large distance, perhaps more than one unit of spatial rapidity. Whereas, if charges are created late in the reaction, the structure of $C'_{ab}$ will be more localized. If the local part of the correlation maintains equilibrium according to lattice gauge theory, the source function for $C'_{ab}$ will have contributions from both early and later charge production. The contributions from the various stages will also depend strongly on the charge indices. For example, $C_{ss}$ will be fed mainly at early times, when most of the strangeness is produced, whereas the majority of the source for $C_{uu}$ will come a later times during hadronization. Investigating the spatial spread of $C'_{ab}$ at the end of the collision provides insight not only into $\chi_{ab}$ at the end of the collision, but into the evolution of $\chi_{ab}$ throughout the event. The actual scales also depend on the diffusion constant and on the initial separation of charges from the first surge \cite{Bass:2000az,Ling:2013ksb}, i.e. decaying flux tubes might pull balancing charges apart as they tunnel through the vacuum in a Schwinger mechanism. Even if the local correlation is not equilibrated, $\chi_{ab}$ can still represent the strength of the local correlation, and might be modeled by some assumption of the non-equilibrium chemical evolution.

An obvious difficulty in extracting $C'_{ab}(\vec{r},\vec{r}',t)$ is that experiments measure only asymptotic momenta. Fortunately, because of the strong collective flow in heavy ion collisions, momenta are strongly correlated with position. A particle's final rapidity $y$ and azimuthal angle $\phi_p$, as defined by their outgoing momenta, are close to the spatial rapidity $\eta_s$ and angle $\phi_r$ describing the last point from which the particles were emitted. Because of thermal motion, the values of $\eta_s$ and $y$ tend to differ by a few tenths of a unit of rapidity \cite{Bass:2000az} and $\phi_p$ and $\phi_r$ differ by a few dozen degrees \cite{Bozek:2004dt}. This smearing out of the correlation can be modeled, but does limit the ability to distinguish correlation features at small length scales.

The measured correlations, known as charge balance functions, are usually defined by the following, or similar, form,
\begin{eqnarray}
\label{eq:Bdef}
B_{h'h}(p'|p)&\equiv&\frac{\langle\Delta\rho_h(p)[\Delta\rho_{h'}(p')-\Delta\rho_{\bar{h}'}(p)]\rangle}
{2\langle\Delta\rho_h(p)\rangle}
-\frac{\langle\Delta\rho_{\bar{h}}(p)[\Delta\rho_{h'}(p')-\Delta\rho_{\bar{h}'}(p)]\rangle}
{2\langle\Delta\rho_{\bar{h}}(p)\rangle}\\
\nonumber
&\approx&\frac{\langle[\Delta\rho_h(p)-\Delta\rho_{\bar{h}}(p)][\Delta\rho_{h'}(p')-\Delta\rho_{\bar{h}'}(p)]\rangle}
{\langle\Delta\rho_h(p)+\Delta\rho_{\bar{h}}(p)\rangle}.
\end{eqnarray}
Here, $h$ refers to some set of hadrons and $\bar{h}$ denotes the corresponding set of anti-particles. For example, $h$ might refer to all positively charged particles and $\bar{h}$ would refer to all the negatively charged ones. The momenta ranges might be such that $p$ refers to any measured track and that $p'$ refers to the relative rapidity. For this case $B_{+-}(\Delta y)$, would represent the probability, given the observation of a track of a given charge, of finding a track of opposite sign vs. the same sign at relative rapidity $\Delta y$. Given that electric charge is conserved, the function $B_{+-}(\Delta y)$ would integrate to unity if the acceptance and efficiency for observing the second particle were perfect. Another example would be $B_{pK^-}$, which would describe the conditional probability of finding a proton vs finding an anti-proton given the observation of a $K^-$ averaged with the conditional probability for finding an anti-proton vs a proton given the observation of a $K^+$. For the limit of zero net charge, a good approximation for LHC energies or the highest RHIC energies, the latter expression in Eq. (\ref{eq:Bdef}) becomes exact and the quantities $\Delta\rho$ can be replaced as $\rho$. The motivation of analyzing experimental charge balance functions, which are functions of some measure of relative momenta such as relative rapidity or azimuthal angle, is to determine or constrain $C'_{ab}$ which are functions of relative position. Assuming one knows the correlations, $C'_{ab}(\vec{r},\vec{r}')$, at the end of the collision, one can determine the correlations of various hadrons as a function of momenta by assuming the differential charges induced by the correlations are distributed thermally \cite{Pratt:2012dz}.

By analyzing charge balance functions indexed by hadronic species from STAR \cite{Wang:2012jua}, it appears that the susceptibilities for strangeness and baryon number grew markedly during early times. This is evidenced by the relatively broad balance functions for $p\bar{p}$ and $K^+K^-$ when plotted as a function of relative rapidity. Not surprisingly, these balance functions are most sensitive to the evolution of $\chi_{ss}$ and the baryon susceptibility, $\chi_{BB}$ \cite{Pratt:2012dz}. In contrast, the observed $\pi^+\pi^-$ charge balance function, which is most sensitive to the electric charge susceptibility, is narrower in central collisions. This is consistent with lattice results, which show that the strangeness and baryon susceptibilities, when scaled for the increasing volume, should stay roughly constant from thermalization until hadronization, while in contrast, the electric charge susceptibility roughly triples as the system expands into the hadronization region. This feeds the correlation $C'$ toward the end of the collision, which results in a narrow peak for the $\pi^+\pi^-$ balance functions. If all the susceptibilities were to evolve similarly with time, the behavior would be opposite. Due to the higher thermal velocities, the $\pi^+\pi^-$ balance function would be the broadest due to the larger thermal velocities for pions due to their relatively small masses. Experimentally, the hierarchy is opposite, with the $p\bar{p}$ balance function being broader than the $K^+K^-$ balance functions, and the $\pi^+\pi^-$ being the narrowest. This behavior was fit with parametric models that assumed the initial formation of a chemically equilibrated QGP, followed by a second surge of charge production consistent with going from QGP susceptibilities to hadronic ones. By treating the diffusive width, in spatial rapidity, from the initial creation of a QGP as one parameter, and the width from the second surge as a second parameter, and then parameterizing the initial QGP susceptibility, it was found that matching the experiment required that the initial quark chemistry was within a few tens of percent of the lattice values \cite{Pratt:2015jsa}. 

Another feature of balance function measurements has been the narrowing of balance functions, indexed only by electric charge, as a function of increasing centrality. This has been observed by STAR at RHIC \cite{Abelev:2010ab,Li:2011zzx,Adams:2003kg,Aggarwal:2010ya}, by NA49 at the SPS \cite{Alt:2004gx}, and by ALICE at the LHC \cite{Abelev:2013csa}. For the most central collisions, the observed widths of the charge balance functions in relative rapidity are consistent with the late surge in charge production mentioned above. For the most peripheral collisions, or for $pp$ collisions, the charge balance functions are broader. The physical cause of these broader correlations is not fully understood. Event generators, like RQMD or URQMD, can match the widths for peripheral collisions \cite{Cheng:2004zy}. But these generators are based on underlying $pp$ event generators\cite{popcorn}, which were simply parameterized to match such widths. The generator RQMD, which like URQMD \cite{urqmd} creates hadrons early, rather than only after a QGP evolves, do not have the narrow feature seen in data, and opposite to the trend seen in data, as the charge balance functions from RQMD broadens with increasing centrality. Charge balance functions have also been measured as a function of beam energy \cite{Alt:2007hk,Adamczyk:2015yga}, a feature that will not be discussed here.

The observed experimental features mentioned above make a case for producing an equilibrated quark-gluon plasma at early times in central collisions of heavy ions, which then lasts for a significant time, perhaps $\gtrsim 5$ fm/$c$, until hadronization. However, these conclusions were made from viewing qualitative trends and by fitting to either a parametric model \cite{Pratt:2015jsa,Pan:2015pzh} or to being unable to fit with a purely hadronic model. The state of the art description of a heavy-ion collision involves modeling the QGP stage with relativistic viscous hydrodynamics \cite{Heinz:2013wva}, then coupling to a hadronic simulation once the temperature falls below $\sim 155$ MeV. The hadronic stage cannot be well described with hydrodynamics because the various species begin to lose thermal contact with one another \cite{sorgepionwind,Pratt:1998gt}. In \cite{Pratt:2017oyf} charge correlations were evolved and the resultant charge balance functions were calculated for a state-of-the-art hydrodynamic model, but hadrons were emitted from the hydrodynamic stage into the vacuum and further evolution in the non-hydrodynamic stage was ignored. In \cite{Pratt:2017oyf} the correlation function $C'$ was seeded in a way that was consistent with local chemical equilibrium. Correlations were evolved according to a diffusion constant for light quarks taken from lattice calculations \cite{Aarts:2014nba}. When the evolution emerged from the hyper-surface and into the hadronic stage, the hadrons, and their charges were created according to thermal arguments. However, the hadrons were then simply emitted into the vacuum where they decayed. Here, a more realistic model is presented, which includes the effects of hadronic rescattering. Such rescattering is is not expected to dramatically alter the results. However, it might not be negligible. For example, if the emission occurs at $T=155$ MeV, many $\rho$ mesons are created. The neutral $\rho$s decay producing balancing $\pi^+\pi^-$ pairs with the invariant mass of the $\rho$. More realistically, such $\rho$s decay and the daughter pions rescatter, altering the structure of the balance function. 

The hadronic simulator B3D \cite{Novak:2013bqa} was employed for the evolution of the hadronic phase here. As in \cite{Pratt:2017oyf} the diffusion of balancing charges was modeled with Monte Carlo methods, which involved tagging correlated pairs. When they were emitted into the vacuum, once could create correlations using only hadrons from the same correlated pair, or from their decay products. This reduced re-combinatoric noise, and made it possible to calculate balance functions at very modest numerical expense. Unfortunately, the complex interactions of the hadron cascade preclude such an efficient treatment. Thus, the more realistic description here carried a significant numerical cost. Hundreds of thousands of cascade events were generated for this analysis. In \cite{Pratt:2017oyf} the results was studied for their sensitivity to several parameters, such as the breakup temperature, the diffusion constant, and the initial charge separation when the QGP was created. Here, results are presented for one default set of parameters, and the discussion is focused on how well the mode reproduces data, and on the effects of the hadronic rescattering. 

The method for modeling the evolution, both in the hydrodynamic and cascade stages, is described in the next section. In addition to describing how the correlations are propagated through the cascade, the method for treating the hydrodynamic stage, as used in Ref. \cite{Pratt:2017oyf} is reviewed. Results for Au+Au collisions with $\sqrt{s_{NN}}=200$ GeV are provided in Sec. \ref{sec:results}. This includes analyses for unidentified particles binned by relative pseudo-rapidity, relative azimuthal angle, and by centrality. Charge balance functions indexed by hadronic species are shown for central collisions, and finally, balance functions  indexed by the angle relative to the reaction plane are presented, an analysis that also provides the $\gamma_p$ correlator related to the chiral magnetic (CME) effect \cite{Adamczyk:2013hsi,Abelev:2009ac}. Each of these calculations are compared to STAR data. The appendix presents a brief description of how a differential balancing charge is translated into hadrons at the hypersurface separating the hydrodynamic and hadronic stages. Prospects for future analysis and measurements are provided in Sec. \ref{sec:summary} along with a summary.

\section{Method}
\label{sec:method}

Charge correlations, $C_{ab}(\vec{r}_1,\vec{r_2})=\langle \rho_a(\vec{r}_1)\rho_b(\vec{r}_2)\rangle$, were propagated through the hydrodynamic stage using the same methods as were applied in \cite{Pratt:2017oyf}. The correlations were based on a hydrodynamic background generated from the iEBE-VISHNU package \cite{Shen:2014vra} using a lattice equation of state \cite{Huovinen:2009yb}. The hydrodynamic treatment provided a description of the stress-energy tensor as a function of the transverse coordinates $x$ and $y$, and the proper time $\tau=\sqrt{t^2-z^2}$. The description assumed boost-invariance along the beam axis, which leads to a translational invariance with respect to the spatial rapidity, $\eta_s=\sinh^{-1}(z/\tau)$. In addition to the proper time $\tau$, the correlations were functions of $x_1,x_2,y_1,y_2$ and the relative spatial rapidity $\Delta\eta_s$, as boost invariance eliminates any dependence on $\eta_{s1}+\eta_{s2}$.  Rather than evolve a five-dimensional quantity, a Monte-Carlo simulation was applied as was performed in \cite{Pratt:2017oyf}. Pairs of sampling particles of charge $q_a,q_b$ were followed through time. The correlation was evolved according to the diffusion equation, and given a source term consistent with maintaining the charge conservation condition in Eq. (\ref{eq:ccons}). Because the correlation was represented by an ensemble of sampling charge pairs, the diffusion equation was not treated as a differential equation, but instead as a random walk with the collision time chosen consistent to be consistent with the diffusion constant. The diffusion coefficient $D$ was a function of the local temperature, and taken from lattice-gauge theory\cite{Aarts:2014nba}. 

As described in the introduction, the correlation was separated into short-range and longer-range pieces,
\begin{eqnarray}
C_{ab}(\vec{r}_1,\vec{r}_2,t)&=&\chi_{ab}(\vec{r}_1,t)\delta(\vec{r}_1-\vec{r}_2)+C'_{ab}(\vec{r}_1,\vec{r}_2,t)
\end{eqnarray}
The delta function is not taken literally, but instead is some function that integrates to unity over a microscopic range describing the equilibrated correlation. For a hadron gas, this would be the size of a hadron, whereas for a QGP, the delta function could practically be literal. If chemistry is equilibrated, $\chi_{ab}(\vec{r},t)$ is indeed the equilibrated charge fluctuation. For this study, it will assumed to be the case, but more generally, one could model the non-equilibrium behavior of the local part of the correlation. Because the local part is accounted for by the single-particle emission from the hydrodynamic stage, only the non-local, or balancing, part needs to simulated.

The non-local part, $C'$ would propagate according to the diffusion equation,
\begin{eqnarray}
\partial_t C'_{ab}(\vec{r}_1,\vec{r}_2,t)&=&
D\langle [\nabla^2\rho_a(\vec{r}_1,t)]\rho_b(\vec{r}_2,t)\rangle
+D\langle \rho_a(\vec{r}_1,t)[\nabla^2\rho_b(\vec{r}_2,t)]\rangle+S_{ab}(\vec{r}_1,t)\delta(\vec{r_1}-\vec{r}_2)\\
\nonumber
&=&D(\nabla_1^2+\nabla_2^2)C_{ab}(\vec{r}_1,\vec{r}_2,t)+S_{ab}(\vec{r}_1,t)\delta(\vec{r_1}-\vec{r}_2),
\end{eqnarray}
where $S_{ab}(\vec{r},t)$ is a source function that feeds the non-local correlation $C'$. Local charge conservation determines the source function,
\begin{eqnarray}
S_{ab}(\vec{r},t)&=&(\partial_t-\nabla\cdot\vec{v}) \chi_{ab}(\vec{r},t).
\end{eqnarray}

The Monte Carlo procedure involved creating sample charge pairs, $q_a,q_b$ at a space-time point $(\vec{r},t)$ with probability,
\begin{eqnarray}
dN_{ab}&=&S_{ab}d^3rdt.
\end{eqnarray}
The first charge $q_a$ was chosen randomly as $\pm 1$, and the second was chosen so the product $q_aq_b$ matches the sign of $dN_{ab}$. Because diffusion describes a random walk, the particles were allowed to move in random directions (in the rest frame of the fluid), with a collision time $\tau_{\rm coll}$ determined by the diffusion equation $\tau_{\rm coll}=6D/v^2$, where $v$ is the velocity between collisions. In each time step $\delta t$, the probability of colliding was $\delta t/\tau$. The velocity was set equal to the speed of light. This approach has two clear advantages compared to solving the differential equation for $C'_{ab}$ described above. First, the Monte Carlo procedure allows one to label balancing pairs, which eliminates combinatoric noise. Secondly, the random walk never violates causality. Invoking causal diffusion in differential equations can also be applied \cite{Kapusta:2014dja,Kapusta:2017hfi,Aziz:2004qu}. Non-diagonal elements of the diffusion constant, or equivalently of the conductivity, were ignored here, which is reasonable for a QGP, but might become questionable if the hydrodynamic description were to be applied for large portions of the hadronic stage.

For an ideal QGP, where the quarks behave independently, the susceptibility, $\chi_{ab}$, is diagonal and the sampled charges effectively represent quark-antiquark pairs created with a rate such that the density of pairs would equal then densities of individual charges. Once off-diagonal correlations exist, as in a hadron gas, $\chi$ becomes more complicated, and the number of pairs is then no more than a Monte Carlo means to represent the correlation function.  

At some point each sample charge passes through the hyper-surface that separates the hydrodynamic and cascade descriptions. In \cite{Pratt:2017oyf}, the hadrons associated with each charge sample were simulated, then correlated with the sample hadrons from the other charge in the pair. For a differential sample charge $d Q_a$ that traverses a hyper-surface element $d\Omega_\mu$, the differential yield for a hadron of species $h$ and charge $q_{h,a}$, whose equilibrium number density is $n_h$, is
\begin{eqnarray}
\label{eq:dNfromdQ}
dN_h&=&n_hq_{h,a}\chi^{-1}_{ab}dQ_b.
\end{eqnarray}
One can see that the average differential charge emitted from the differential hadron yield $dN_h$ is indeed $dQ$. Summing over the hadron species, the net charge carried by the hadrons, $dQ'$, is
\begin{eqnarray}
dQ'_a&=&\sum_h q_{h,a} dN_h \\
\nonumber
&=&\sum_h q_{h,a}n_hq_{h,b}\chi^{-1}_{bc}dQ_c\\
\nonumber
&=&\chi_{ab}\chi^{-1}_{bc}dQ_c\\
\nonumber
&=&dQ_a.
\end{eqnarray}

The momentum of the particle is then chosen according to the Cooper-Frye formula\cite{Cooper:1974mv},
\begin{eqnarray}
\label{eq:CF}
dN&=&\frac{d^3p}{E_p}~p\cdot d\Omega~f(\vec{p}),
\end{eqnarray}
where $f(\vec{p})$ is the phase space density at the point on the hyper-surface element $d\Omega$. The phase space density in this instance is the thermal form determined by the local temperature, collective velocity and the viscous corrections to the stress-energy tensor. Implementing the Cooper-Frye formula is complicated by the fact that a small portion of the sampled phase space has a negative contribution, where $p\cdot d\Omega$ is negative, which cannot be easily represented in a Monte Carlo representation. This only occurs for space-like hyper-surface elements, which provide a small fraction of the overall emisssion. A variety of strategies have been implemented to account for these negative contributions \cite{Huovinen:2012is,Becattini:2012sq,Becattini:2012xb,Pratt:2014vja} or to account for viscosity-related anisotropies of the momentum distribution \cite{Pratt:2010jt}. In the appendix \ref{sec:CFappendix} we describe the approximation used here that has the advantage of perfectly satisfying charge conservation. 

In \cite{Pratt:2017oyf}, where there was no hadronic cascade, the charge balance functions were divided into two contributions, denoted 1A and 2A below:
\begin{itemize}
\item[1A.] Correlations from the hydrodynamic stage were projected into the final state. As stated above, these correlations were represented in a Monte Carlo procedure by sampling charge pairs $q_1$ and $q_2$. These charges, which could be $\pm u,\pm d$ or $\pm s$, each carried the information of the hyper-surface element through which it left the hydrodynamic stage and entered the vacuum. Each charge produced hadrons, via a Monte Carlo procedure according to the weights described above. An additional multiplicative factor for producing hadrons was added to increase the numerical efficiency of the procedure. The hadrons were decayed, with the decayed hadrons assigned to the stream from which the decaying hadrons originated. Hadrons from a stream originating from a specific sample charge, were only correlated with those hadrons from the stream coming from the paired charge. No correlations were considered from hadrons coming from the same charge, or from hadrons coming from two charges that were not in the same pair. By not mixing in hadrons from uncorrelated pairs, combinatoric noise was largely avoided. Through this procedure, the numerators to the balance function represent the correlation that existed in the hydrodynamic stage, but ignored any evolution of charge correlations that might be generated after the hydrodynamic stage, including those from decays.
\item[2A.] Correlations from decays were generated by first sampling the hyper-surface, ignoring the sampling charge pairs described above. Such decays can account for over 40\% of the charge balance function's overall normalization \cite{Bozek:2003qi}. These uncorrelated hadrons were then decayed, and the charge balance functions were incremented only from hadrons coming from the same decay chain. Again, combinatoric noise was avoided because only those hadrons with the same ancestor were correlated. The hadrons from this procedure were also used to generate the denominator of the charge balance function.
\end{itemize}
Because the balance function numerators and denominators are all represented by a Monte Carlo sampling, accounting for the experimental acceptance and efficiency was rather straightforward. Because the hydrodynamic calculations assumed longitudinal boost invariance, the correlated hadron pairs were randomly boosted by a rapidity $\Delta y$ so that the first hadron would have a rapidity randomly between $\pm 1$. Because both hadrons were boosted by the same rapidity, this did not change the correlation. Hadrons were then weighted by the experimental efficiency before the contributions were used to increment the balance functions. For identified particles, a sophisticated routine was applied that returns the efficiency as a function of pseudo-rapidity and transverse momentum \cite{WestfallAcceptance}. For the balance functions for non-identified charged particles, a very simple routine was applied. For this simple routine, all particles with pseudo-rapidities between $\pm 1$ and transverse momenta between 200 MeV/$c$ and 2 Gev/$c$ were accepted and assigned a uniform efficiency. Various over-sampling rates from the Monte Carlo procedures were also applied to calculations of both the balance function numerators and denominators. As a test of the procedure, calculations were performed with perfect acceptance and efficiency. In that case, the charge balance function for unidentified charged particles should integrate to unity. In practice, due to the numerical accuracy of the representation of the hyper-surface and finite hydrodynamic resolution, the balance function integrated to within a few tenths of a percent of unity.

For this study, a hadronic cascade was added to model the post-hydrodynamic stage. Both steps of the procedure (1A and 2A above) of \cite{Pratt:2017oyf} were modified:
\begin{itemize}
\item[1B.] Hadrons generated from the sample charges were propagated through a cascade. The cascade evolved two sets of particles. The first set was one of uncorrelated particles generated from the hyper-surface elements consistent with the single-particle phase space density. This first set is used as a base for scattering the second class of particles. The second set were those hadrons generated from the pairs of sample charges representing the correlation function of the hydrodynamic stage. These are the same as those particles from (1A) described above. As was done without the cascade, these hadrons are labeled by the sampling charge responsible for their emission. Unstable hadrons were allowed to decay, with these labels being passed on to their decay products. Additionally, these hadrons were allowed to elastically scatter from those of the first set, but not with those of the second set. If hadron $h_2$, from the second set, scattered off $h_1$ from the first set, only $h_2$ had its trajectory altered. These scatterings provide a approximate way to model the evolution of the charge coming from the second particle. By ignoring resonant interactions (aside from decays), and by fixing the cross sections independent of isospin, the effects of the scatterings is fully represented by the altered trajectories of the $h_2$ hadrons. Because resonant interactions represent a good fraction of the scatterings in a hadronic gas, a larger elastic cross section of 20 mb was assumed for the scattering. As was done in the (1A) for the previous study, incrementing balance function numerators involved only pairing hadrons coming from the same correlated charge pair, but not from the same sampling charge .
\item[2B.] As was done in (2A) above, an uncorrelated set of hadrons was emitted from the hyper-surface. However, in this case the particles were allowed to fully interact, including resonant recombinations and decays. Because such interactions mix and share charges in non-trivial ways, tagging could not be used to identify a pair of hadrons as coming from the same original source. Thus, all pairs of hadrons were considered when constructing the numerator of the balance function, similarly to how experimental data is considered. For this method combinatoric noise was overcome by increasing the number of events. For most centralities 80,000 events were analyzed, each covering $\pm$ 5 units of rapidity. Combined with the fact that particles are not lost due to efficiency, the noise is similar to what one would expect for experimental analyses if a million events were recorded for a given centrality.
\end{itemize}
Correlations generated in the cascade, i.e. those described in (2B) would seem to be well modeled with this procedure. The method is somewhat numerically intensive, but the cascade B3D \cite{Novak:2013bqa} propagates several events per second, which makes the procedure quite tenable. The treatment of correlations from the hydrodynamic stage seems less satisfactory due to the scatterings being only elastic. However, because such correlations involve coupling hadrons from two different streams, the main goal is to understand how the spread of the charge carried by a hadron spreads out in the cascade. This spread involves balancing the effect of diffusion, which spreads out the charge, and cooling which more focuses the charge. Neither of these effects during the cascade stage is significant, and little change is noticed between (1A) and (1B). The effect of the cascade from (2B) vs (2A) is potentially noticeable. For example, during the cascade particles decay, and their products re-scatter. At the end of the reaction, relatively few heavy resonances, like the $\rho^{(0)}$, remain. Thus, there are many more $\pi^+,\pi^-$ pairs with the invariant mass of the $\rho$ for method (2A) than for (2B), where the re-scattering was considered. In (2B), the products of a $\rho$ decay would still contribute to the balance function, but after re-scattering, their invariant mass distribution would be different. Such effects are subtle when viewing the balance function in relative rapidity or relative azimuthal angle, but would be pronounced for balance functions binned by invariant mass.

The most important missing feature in this treatment is probably that baryon annihilation is ignored. Because annihilation and regeneration of baryons should be performed consistently, and because regeneration can be somewhat numerically costly, it was neglected in this treatment. Annihilation might reduce the baryon yields by 25\% or more \cite{Pan:2014caa,Steinheimer:2017vju,Steinheimer:2012rd,Steinheimer:2012bn}, which should provide a significant dip in the $p\bar{p}$ balance functions at small relative momenta, relative rapidity or relative azimuthal angle. This improvement is a priority for the next study.

\section{Results}
\label{sec:results}

Here, model calculations are compared to measurements of the STAR Collaboration at RHIC for Au + Au collisions at $\sqrt{s}_{\rm NN}=200$ GeV. The next subsection presents results for balance functions of non-identified particles, binned by relative pseudo-rapidity and relative azimuthal angle. Calculations are shown for several centralities. The following subsection shows results for balance functions indexed by hadron species. All combinations of pions, kaons and protons are calculated, and compared to data for $\pi\pi$, $p\bar{p}$, $KK$, and $pK$. The final subsection compares balance functions of unidentified particles as a function of relative azimuthal angle, and also binned by the direction of the first pion relative to the reaction plane. This provides a detailed test of collective flow and also provides insight into the correlation measure, $\gamma_p$, which has been proposed as a signal of the chiral magnetic effect.

\subsection{Balance Functions for Unidentified Hadrons}

First, the model produced charge balance functions for unidentified hadrons, i.e. all charged particles without discrimination based on species but distinguished by whether their charges were positive or negative. For perfect acceptance and efficiency, such balance functions would integrate to unity, because for each positive particle, there exists one additional negative charge relative to the positive. This additional charge can be accounted for by some combination of additional negatives or a reduced number of positive tracks. In practice, the experimental balance functions integrate to approximately 0.35. This shortcoming comes from a combination of efficiency and acceptance. The normalization is first reduced by the efficiency, which varies from approximately 0.7 for central collisions to approximately 0.8 for more peripheral collisions. Because the STAR measurement is limited to tracks with a finite range of pseudo-rapidities, $-1<\eta<1$, a significant fraction of the balancing charge, for an observed charge, falls outside the rapidity range. Finally, the measurement is confined to particles with transverse momenta, $200$ MeV $<p_{\rm t}<2$ GeV, and to particles with a DCA (distance of closest approach) of 3.0 cm or less. This latter cut reduces some of the contributions from weak decays. Thus, because there is only a $\approx 50\%$ chance of a balancing charge being in the acceptance, and because the imperfect efficiency reduces the chance of observing a charge in the acceptance by $\approx 75\%$, the balance functions for unidentified particles integrate to $\approx 0.35$. The calculations shown here for unidentified particles applied a crude acceptance and efficiency filter. The acceptance cuts matched those of the experiment, but the representation of the efficiency was approximate. Here, the efficiency was assumed to be independent of transverse momentum, whereas in reality it has a modest dip for the low $p_{\rm t}$ range. The constant efficiency was chosen to be 0.7 for $0-5\%$ centrality collisions, and $0.72,0.74, 0.76, 0.78$ and $0.8$ for centralities of $10-20,20-30,30-40,40-50$ and $50-60\%$ respectively. A more accurate representation of the efficiency might alter results by a few percent.

\begin{figure}
\centerline{\includegraphics[width=0.5\textwidth]{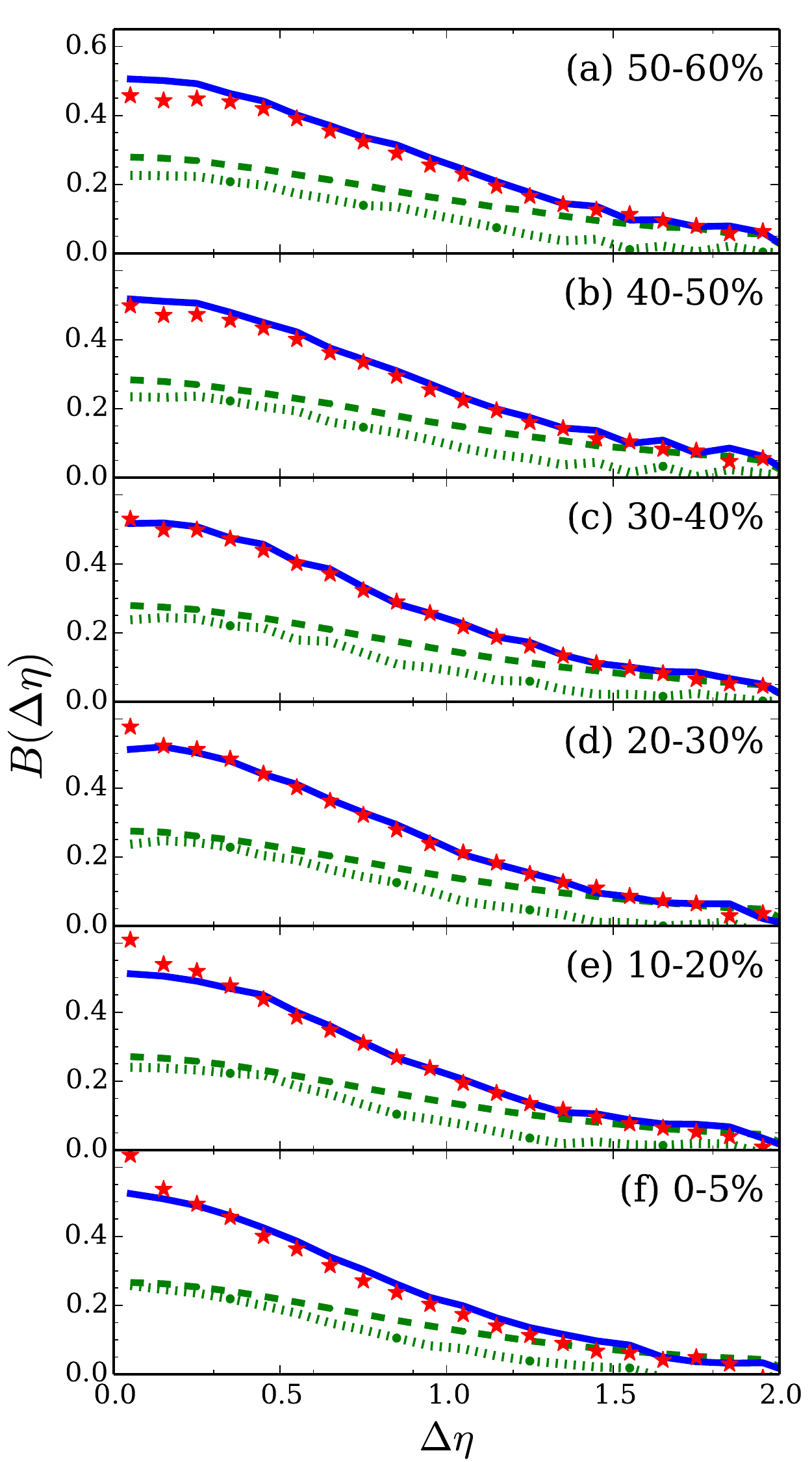}
\includegraphics[width=0.5\textwidth]{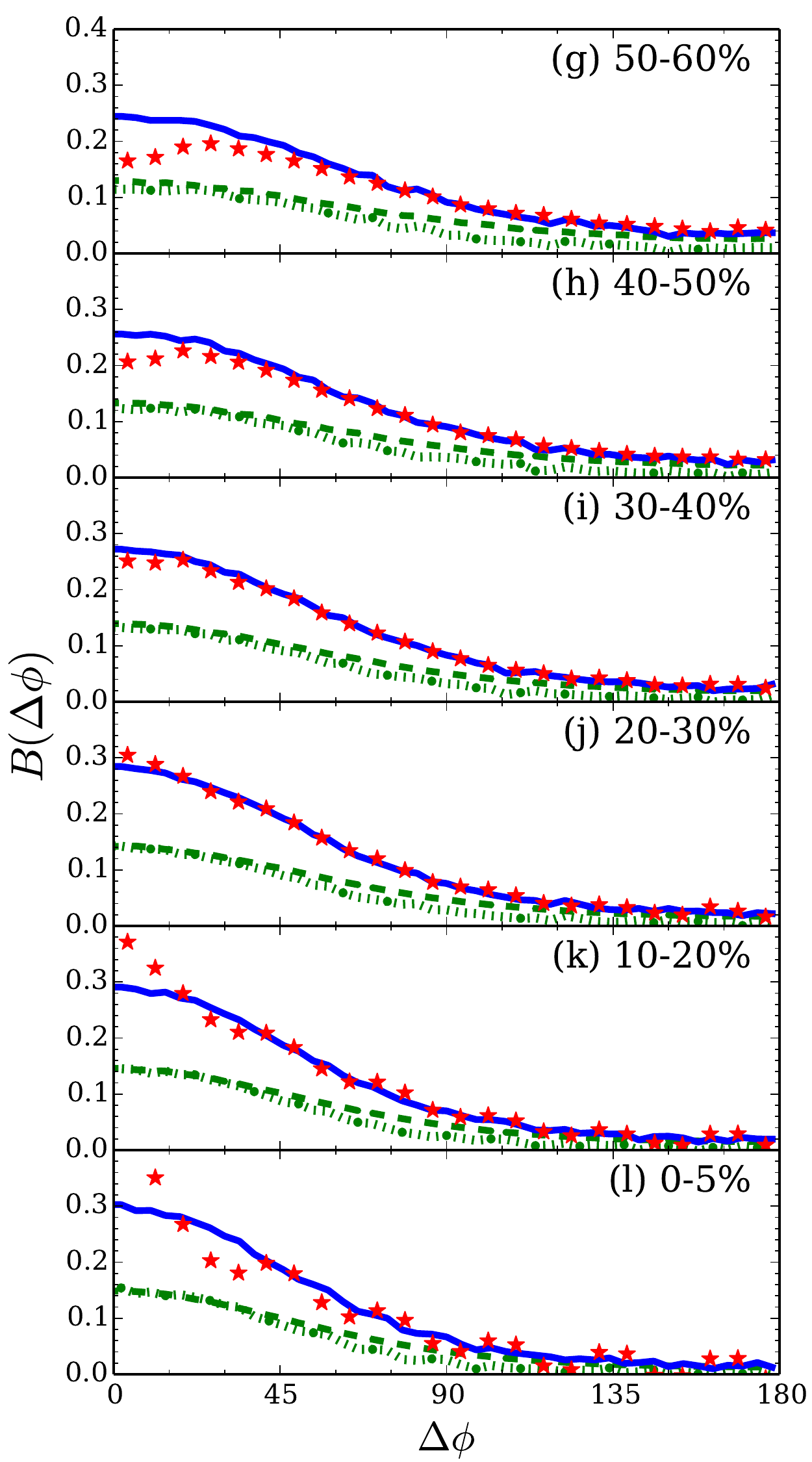}}
\caption{\label{fig:bf_allcents}(color online)
Panels (a-f): Charge balance functions for unidentified charged particles binned by relative pseudo-rapidity for six different centralities, from 0-5\% to 50-60\%. The model (solid blue lines) approximately reproduces the narrowing of the experimental balance functions (stars) with increasing centrality.\\Panels (g-l): The same as (a-f), but binned by relative azimuthal angle. The larger volumes for more central collisions make it more difficult for charges to diffuse to regions with different radial flow, hence the balance functions are narrower. The contributions from the hydrodynamic correlations (red dashed lines) and from the correlations that originated in the cascade (green dotted lines) are of similar strength, with the cascade contribution being narrower. Oscillations of the experimental balance functions for the most central collisions in panels $k$ and $l$ are likely from the sector boundaries of the STAR experiment. Some of the deviations for small $\Delta\eta$ and $\Delta\phi$ might be due to femtoscopic correlations.
}
\end{figure}
Figure \ref{fig:bf_allcents} shows charge balance functions binned by relative pseudo-rapidity, $\Delta\eta$, and relative azimuthal angle, $\Delta\phi$, for several centrality bins. Pseudo-rapidities, $\eta$, are approximate surrogates for the rapidities $y$. Defined in terms of the polar angle relative to the beam axis,
\begin{eqnarray}
\eta&=&\frac{1}{2}\ln\left(\frac{1+\cos\theta}{1-\cos\theta}\right),
\end{eqnarray}
they are equal to the rapidity in the limit that the particles move at the speed of light and $\cos\theta=v_z$. For unidentified particles, the mass, and therefore the velocities are unknown, hence analyses are performed for relative pseudo-rapidity rather than for relative rapidity. For more central collisions, the model produced the experimental data remarkably well, with the exception being the most central azimuthal correlations. The experimental measurements binned by $\Delta\phi$ are affected by the sector boundaries of the STAR Time Projection Chamber. These boundaries result in acceptances that depend on azimuthal angle, which differ for positive and negative tracks because the tracks curve in opposite directions from the longitudinal magnetic field. Because of the curvatures, the positions of the dips in acceptance are displaced from the angles of sector boundaries. This displacement is opposite for oppositely charged particles, which results in structures in the balance function when binned by $\Delta\phi$. These correlations affect all pairs of particles, not just the correlation between a single track and a balancing pair, hence their strength relative to the true correlation increases with multiplicity. In the experimental analysis, a balance function for mixed events was constructed, and subtracted from the same-event distributions. This procedure very much reduced the magnitude of these oscillations, but did not make them completely disappear. Aside from the oscillation, the agreement of the model to data is remarkable.

To correct for the finite acceptance in relative rapidity, both the experimental and model calculations in the left-side panel of Fig. \ref{fig:bf_allcents} were divided by a factor $(1.0-\Delta\eta/2)$ \cite{Pratt:2003gh}. Discrepancies for the first bin, whether in relative rapidity or relative azimuthal angle, can be caused by femtoscopic correlations or track merging, and should not be given much consideration here. For more peripheral collisions, femtoscopic correlations can extend to larger relative momentum due to the smaller source sizes, and might distort the first few bins. 

Results from both the model and from the experimental analysis show a narrowing of the balance functions with increasing centrality, qualitatively consistent with the predictions of \cite{Bass:2000az}. The narrowing in the data appears slightly more pronounced than in the model. The stronger narrowing for central collisions, could be caused by stronger collective flow for those reactions, or perhaps by a reduced contribution from resonances, should the more peripheral collisions not reach the same degree of chemical equilibration. In \cite{Bass:2000az}, the narrowing was expected to come from the delayed hadronization associated with a longer-lived QGP state. Because the majority of electric charge is created at or near hadronization, these balancing charges would have less chance to separate if they were produced after the system had expanded and the velocity gradients subsided somewhat. However, it is difficult to pinpoint the exact causes of the narrowing. Species-dependent balance functions, which are the subject of the next subsection, provide a superior means for identifying delayed hadronization.

\subsection{Balance Functions Indexed by Hadronic Species}

Balance functions of the type,
\begin{eqnarray}
B_{h'h}(p'|p)&=&\frac{\langle(\rho_h(p)-\rho_{\bar{h}}(p))(\rho_{h'}(p')-\rho_{\bar{h}'}(p'))\rangle}{\langle\rho_h(p)\rangle+\langle\rho_{\bar{h}}(p)\rangle},
\end{eqnarray}
where $h$ and $h'$ refer to specific hadronic species, provide the means to disentangle the three-by-three correlation matrix $C_{ab}(r-r')$ in coordinate space. Again, the momenta $p$ will typically be any observed particle, while $p'$ will refer to the relative rapidity or relative azimuthal angle. Here, we consider the species as pions, kaons or protons, which thus provides six independent combinations of possible species-dependent balance functions: $B_{\pi^+,\pi^-},B_{K^+K^-},B_{p\bar{p}},B_{K^-\pi^+},B_{\bar{p}\pi^+}$ and $B_{\bar{p}K^+}$. By symmetry, the numerators for $B_{\bar{p}K^+}$ and $B_{pK^-}$ are identical, and opposite to $B_{\bar{p}K^+}$ and $B_{pK^+}$.  The correlation matrix $C_{ab}(\vec{r}-\vec{r}')$ is symmetric, and because of isospin symmetry between the $u$ and $d$ quarks, has only 4 independent elements, $C_{uu}=C_{dd}$, $C_{ud}$, $C_{us}=C_{ds}$ and $C_{ss}$.

\begin{figure}
\centerline{\includegraphics[width=0.45\textwidth]{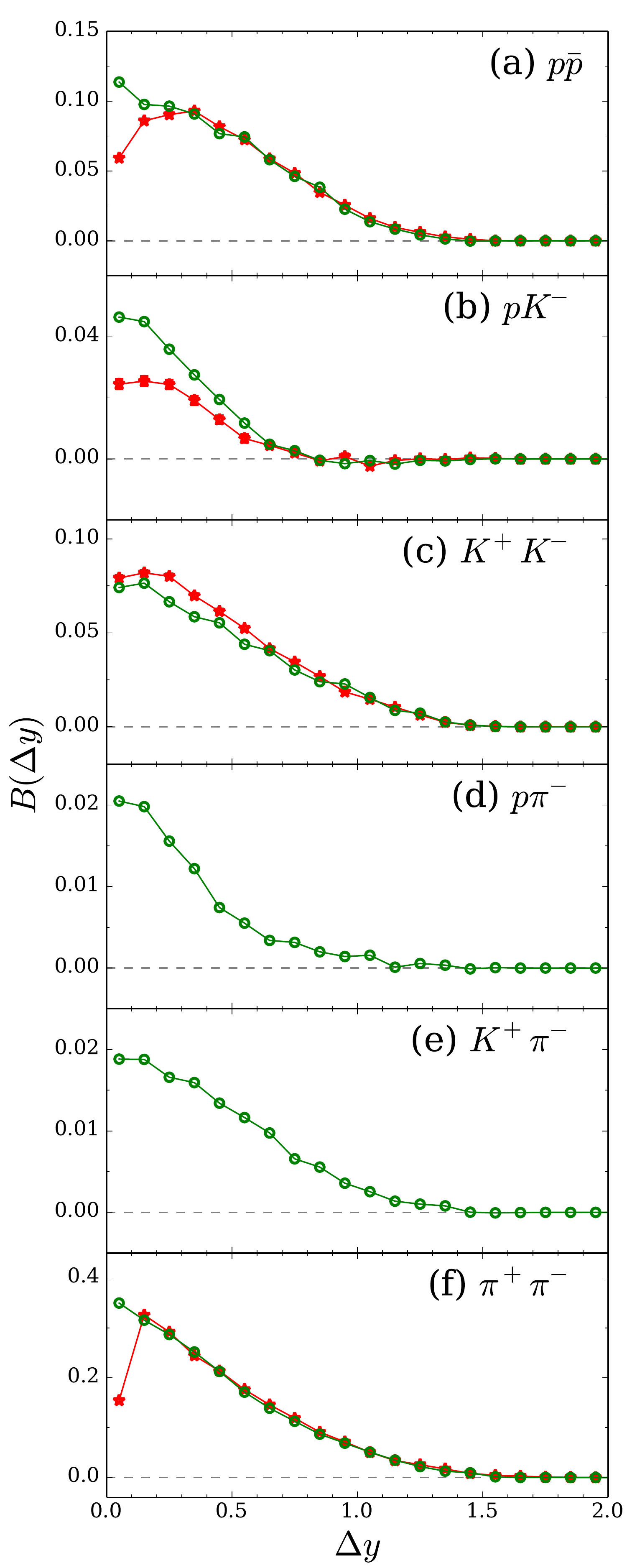}}
\caption{\label{fig:general_y_rhic_cent0_5}(color online)
Balance functions, indexed by hadronic species and binned by relative rapidity, are shown for central (0-5\%) Au+Au collisions at $\sqrt{s}_{NN}=200$ GeV. The model calculations (blue lines) are compared to preliminary measurements from the STAR Collaboration at RHIC \cite{Wang:2012jua} (red stars).  Matching the relatively broader structure of the $K^+K^-$ and $p\bar{p}$ balance functions relative to the $\pi^+\pi+-$ balance function provides compelling evidence that the a chemically equilibrated quark-gluon plasma was created.  Unfortunately, such conclusions are tempered by the failure of the model to reproduce the $pK^-$ experimental balance functions. 
}
\end{figure}

The resolving power of this set of correlations for determining $C_{ab}$, described in Sec. \ref{sec:intro}, is due to the varying quark content of the various hadrons. For example, $K^+K^-$ balance functions are strongly influenced by the $ss$ component of the charge correlation. Figure \ref{fig:general_y_rhic_cent0_5} presents calculations for all six combinations one can make with pion, protons and kaons. Balance functions could be constructed with other species, such as lambdas or neutrons, but technical issues make such measurements difficult. The acceptance for the model calculations mirrored what was applied in the STAR analyses. Transverse momenta were confined to $200$ MeV/$c < p_{\rm t}<1.6$ GeV/$c$, rapidity cuts of $-0.9<y<0.9$, and a DCA cut of 3.0 cm were applied. Additionally, a sophisticated filter provided by STAR was applied to the model to reproduce the effects of STAR's efficiency \cite{WestfallAcceptance}. For $K^+K^-$ pairs, an additional invariant mass cut was applied to eliminate the contributions of neutral kaon and phi meson decays.

In the calculations, because chemical equilibrium is assumed for the hydrodynamic stage, strange quark production is mainly confined to the early stages of the collision when the QGP is formed. This contrasts to the production of electric charge, where most occurs at or near hadronization. Given that the $K^+K^-$ balance function is mainly driven by $ss$ correlations and that the $\pi\pi$ balance functions are mainly driven by the correlations of electric charge, the fact that the kaon correlations are broader than the pion correlations, and that these widths are rather well reproduced by the model, makes a good case that central collisions at RHIC produce what is close to a chemically equilibrated quark-gluon plasma, and that equilibration occured at early times. The $p\bar{p}$ balance functions further strengthens this claim. For a chemically equilibrated system, the baryon susceptibility changes little during hadronization, which results in baryon-baryon correlations being driven by early charge creation. Again, the model quantitatively reproduces the widths of both the proton and pion balance functions, with the proton balance function being broader. The modest discrepancy of the $K^+K^-$ balance functions might be corrected by choosing a slightly lower initial width, $\sigma_0$, for the balancing charges at the formation time of the QGP, $\tau_0=0.6$ fm/$c$. This sensitivity of the $K^+K^-$ balance function to $\sigma_0$ was shown in \cite{Pratt:2017oyf}. From those results, it would seem that reducing $\sigma_0$ from the value of 0.75 assumed here to $\approx 0.6$ might make up such a difference while changing other results rather little. The small dip at low relative rapidity in the experimental $p\bar{p}$ balance function was not reproduced in the model. But this discrepancy was expected given the lack of baryon annihilation in the cascade calculations, a correction planned for future studies.

The promising reproduction of the $\pi^+\pi^-$, $K^+K^-$ and $p\bar{p}$ balance functions is tempered by the failure to reproduce the $pK^-$ balance function. In this case the model calculations are approximately 75\% higher than the experimental balance functions. This same discrepancy was seen in \cite{Pratt:2017oyf}, and none of the variations performed that study seemed particularly strong enough to bring model calculations in line with the experimental result. At this time, the experimental results are preliminary, and only appear in a thesis.

Some guidance in resolving the discrepancies in the $pK^-$ balance function can be obtained by studying the hadronic source functions presented in Fig.~6 of Ref.\cite{Pratt:2017oyf}, which reveal that the $pK^-$ correlations (panel c) receive much larger relative contributions from intermediate stages of the system's evolution than do the correlations of other hadronic species (panels b, d-g).  The $pK^-$ balance function consequently reflects the subsequent evolution of charge pairs in a way which is somewhat less sensitive than the other balance functions to the initial stage of the collision.  This means that uncertainties in the later stages of the collision evolution - including the cascade phase - will tend to dominate the $pK$ balance function without significantly altering the rest of the balance functions.  Thus, it is possible that better accounting for the effects of flavor-dependent freeze-out and improving the description of $\chi_{h h'}$ at late times could reduce some of the discrepancies currently seen in the $pK^-$ charge balance functions without sacrificing the agreement in the remaining correlations or compromising the inference to the production of a chemically equilibrated QGP in the early stages of the collision.

\subsection{Balance Functions Binned by Angle Relative to the Reaction Plane}

The width of the balance function in azimuthal angle is mainly determined by two factors: the relative separation in coordinate space of the balancing charges and the strength of the radial collective flow. Charges that are close to one another in coordinate space will also likely be emitted with similar velocities because they would come from regions with similar collective flow. More central collisions have lower breakup temperatures and higher collective flow velocities. Furthermore, the larger sizes make it more difficult to diffuse balancing charges to regions with different collective velocities. Thus, balance functions binned by relative azimuthal angle tend to be significantly narrower in more central collisions, a trend seen in both the data and models in Fig. \ref{fig:bf_allcents}.

For mid-central heavy-ion collisions, elliptic flow develops from anisotropic pressure gradients caused by the initial elliptic transverse spatial anisotropy of the participant region. Because of the higher flow velocity in the reaction plane, the balance function should be narrower for in-plane vs. out-of-plane pairs. Here, reaction-plane-dependent balance functions are defined as
\begin{eqnarray}
B(\Delta\phi|\phi_1\in\Phi)&=&\frac{N_{+-}(\phi_1\in \Phi,\phi_1+\Delta\phi)+N_{-+}(\phi_1\in\Phi,\phi_1+\Delta\phi)}
{N_+(\phi\in\Phi)+N_-(\phi\in\Phi)}\\
\nonumber
&-&\frac{N_{++}(\phi_1\in \Phi,\phi_1+\Delta\phi)-N_{--}(\phi_1\in\Phi,\phi_1+\Delta\phi)}
{N_+(\phi\in\Phi)+N_-(\phi\in\Phi)},
\end{eqnarray}
where the first charge is required to be in some window $\Phi$. Three windows were evaluated: $0^\circ<\phi_1<7.5^\circ$, $37.5^\circ<\phi_1<52.5^\circ$, and $82.5^\circ<\phi_1<90^\circ$, where an angle $\phi_1=0$ refers to the reaction plane. For charges outside the first quadrant, $0<\phi<90^\circ$, momenta were reflected about the reaction plane and/or the $x=0$ plane to exploit the reflective symmetries. 

\begin{figure}
\centerline{\includegraphics[width=0.45\textwidth]{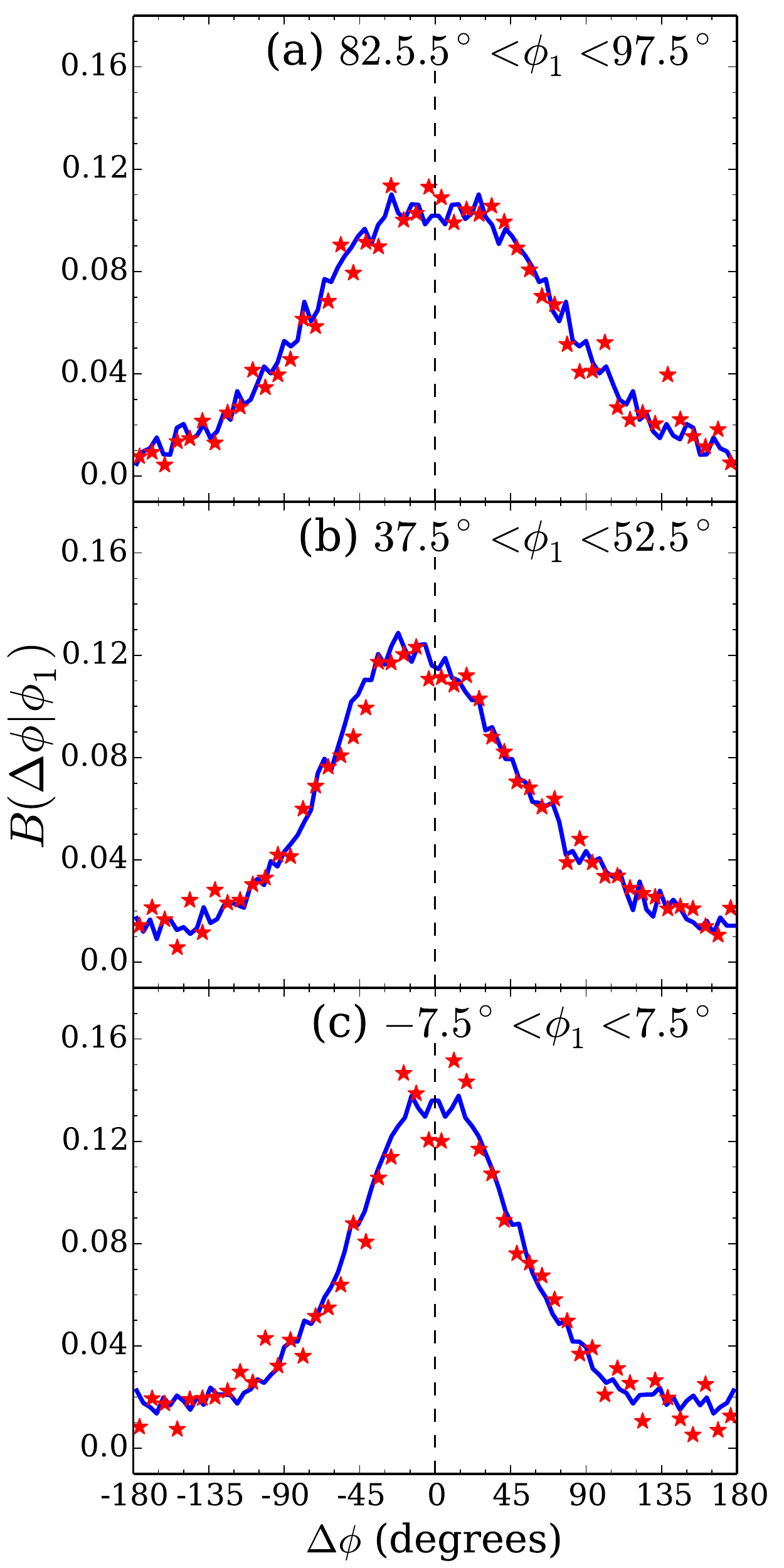}}
\caption{\label{fig:bf_phi1phi_star}Balance functions plotted as a function of relative azimuthal angle are additionally constrained by the angle of the first charge, $\phi_1$, which is measured relative to the reaction plane. The in-plane balance function, $\phi_1\approx 0$, is significantly narrower than the the out-of-plane balance function, $\phi_1\approx 90^\circ$, due to the stronger collective flow. When $\phi_1\approx 45^\circ$ the balance function skew towards negative $\Delta\phi$ because the balancing charge is more likely to be found closer to the reaction plane, where more particles are emitted. The model calculations (blue lines) have been scaled by a factor of 0.94 to match the normalization of the preliminary experimental results from STAR \cite{Wang:2012jua} (red stars). After adjusting the normalization the experimental and model results are in remarkable agreement.}
\end{figure}
Figure \ref{fig:bf_phi1phi_star} displays calculations alongside results from STAR for events in the 40-50\% centrality class. As expected, the in-plane balance functions, $\phi_1\approx 0$, are narrower than the out-of-plane balance functions, $\phi_1\approx 90^\circ$. The difference is striking, and underscores the strength of elliptic flow at these energies. Balance functions with $\phi_1\approx 45^\circ$ are also presented. In this case, seeing a charge near $45^\circ$, more strongly enhances the probability of finding a balancing charge for negative $\Delta\phi$ than for positive $\Delta\phi$. This is expected because there are more charges for $\phi_2\lesssim 45^\circ$ than for $\phi_2\gtrsim 45$ because of elliptic flow. The model calculations in Fig. \ref{fig:bf_phi1phi_star} were all scaled down by a factor of 0.94 so that the experimental and model balance functions would have very similar normalizations. After the normalization was taken into account, the experimental and model calculations were in remarkably good agreement for all three cuts on $\phi_1$. 

As a signal of the chiral magnetic effect (CME), the observable $\gamma_p$ was proposed \cite{Adamczyk:2013hsi,Abelev:2009ac},
\begin{eqnarray}
\gamma_p&=&\langle \cos(\phi_1+\phi_2)\rangle_{os}-\langle\cos(\phi_1+\phi_2)\rangle_{ss},
\end{eqnarray}
where $os$ and $ss$ refer to opposite-sign and same-sign respectively, with the angles being measured relative to the reaction plane. The observable was designed to find evidence of the coherent magnetic fields from the spectator portions of the colliding nuclei to rotate into electric fields due to the coupling between $\vec{E}\cdot\vec{B}$ in the electromagnetic sector to the anomalous charge density in the QCD sector, $\vec{E}_{a,{\rm QCD}}\cdot\vec{B}_{a,{\rm QCD}}$. Strong longitudinal color electromagnetic fields are expected to exist in the early stages which lead to strong anomalous charge densities. These fluctuate in sign, from one flux tube to another, but given that multiple charges might originate from a single flux tube, one might expect some effects of coherence. The coherence of these fields, combined with the coherence of the external magnetic field should serve as a source for a generated electric field. This field would be randomly parallel or anti-parallel to the original magnetic field, and would lead to a small correlation between same-sign charges out of the reaction plane. Observing the CME effect would represent a landmark achievement as it would represent the first observation of coupling between the anomalous charge densities in the electromagnetic and QCD sectors. Unfortunately, the effect might be too small to be observed given (1) the effect would involve a power of the fine structure constant, (2) the magnetic fields may dissipate before there is enough charge to generate a current, and (3) there should be many domains with random anomalous charge densities in the QCD sector. The main background for this observable is the combination of local charge conservation imprinted onto elliptic flow responsible for the correlations in Fig. \ref{fig:bf_phi1phi_star} \cite{Schlichting:2010qia}. Here, we present results for $\gamma_p$ from this model and compare to STAR results to see the degree to which this background explains STAR's result.

Rewriting the observable using angle addition formulas as
\begin{eqnarray}
\gamma_p&=&\langle\sin\phi_1\sin\phi_2\rangle_{ss}-\langle\sin\phi_1\sin\phi_2\rangle_{ss}
+\langle \cos\phi_1\cos\phi_2\rangle_{os}-\langle\sin\phi_1\sin\phi_2\rangle_{os},
\end{eqnarray}
one can see that an enhancement of same-sign pairs out-of-plane ($|\sin\phi|\approx 1$) leads to a positive value for $\gamma_p$. A positive value for this moment can also be caused by an enhancement of opposite-sign pairs in-plane, which is precisely what is seen in Fig. \ref{fig:bf_phi1phi_star}). To better illustrate how charge balance superimposed on elliptic flow could affect $\gamma_p$, one can rewrite $\gamma_p$, again using angle addition formulas, as
\begin{eqnarray}
\label{eq:gp3}
\gamma_p&=&\left\{\langle\cos 2\phi\rangle\langle\cos\Delta\phi\rangle_{ss}
-\langle\cos 2\phi_1\rangle\langle\cos\Delta\phi\rangle_{os}\right\}\\
\nonumber
&+&\left\{
\left(\langle\cos 2\phi_1\rangle\langle\cos\Delta\phi\rangle_{ss}-\langle\cos 2\phi_1\rangle\langle\cos\Delta\phi\rangle_{ss}\right)
\left(\langle\cos 2\phi_1\rangle\langle\cos\Delta\phi\rangle_{os}-\langle\cos 2\phi_1\rangle\langle\cos\Delta\phi\rangle_{os}\right)\right\}\\
\nonumber
&-&\left\{
\langle\sin 2\phi_1\sin\Delta\phi\rangle_{ss}+\langle\sin 2\phi_1\sin\Delta\phi\rangle_{os}
\right\}\\
\nonumber
&=&\frac{1}{2\pi (dN_{\rm ch}/d\eta)}\left\{v_2\int d\Delta\phi~B(\Delta\phi)\cos(\Delta\phi)\right.\\
\nonumber
&+&\left.\frac{1}{2\pi}\int d\phi_1d\Delta\phi~B(\Delta\phi|\phi_1)\cos(2\phi_1)\cos(\Delta\phi)
  -\frac{1}{2\pi}\int d\phi_1d\Delta\phi~B(\Delta\phi|\phi_1)\sin(2\phi_1)\sin(\Delta\phi)
\right\}.
\end{eqnarray}
Aside from the prefactor, the first term in Eq. (\ref{eq:gp3}) represents the elliptic flow $v_2=\langle\cos 2\phi\rangle$, multiplied by the average $\cos \Delta\phi$ of the balance function, which is a measure of its narrowness. The second term represents a correlation between $\cos 2\phi_1$ and $\cos\Delta\phi$, or a correlation between $\cos 2\phi_1$ and the narrowness of the balance function. For reaction-plane balance functions that are narrower for $\phi_1\approx 0$ than for $\phi_1\approx 90^\circ$, this correlation is positive. Indeed, a positive correlation can be seen by comparing the $\phi_1\approx 0 $ and $\phi_1\approx 90^\circ$ balance functions in Fig. (\ref{fig:bf_phi1phi_star}). Finally, the final term represents the correlation between $\sin 2\phi_1$ and $\sin\Delta\phi$. Given that $\sin 2\phi_1$ is largest for $\phi_1\approx 45^\circ$, inspection of Fig. \ref{fig:bf_phi1phi_star} shows that this contribution is also positive. Each of these three contributions is positive and of similar magnitude. Thus, a calculation of reaction-plane-dependent charge balance functions also provides also a calculation of $\gamma_p$.

\begin{figure}
\centerline{\includegraphics[width=0.5\textwidth]{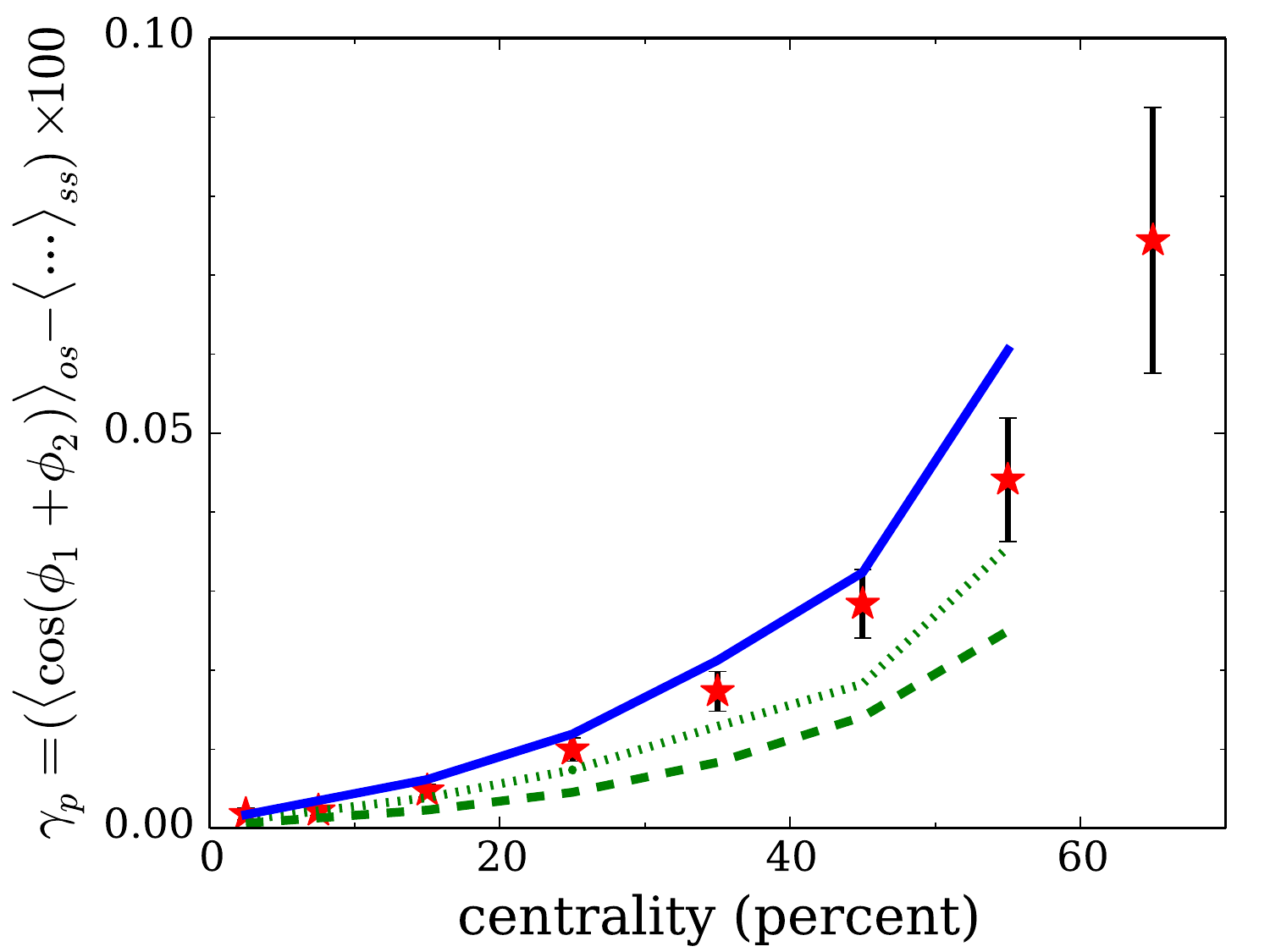}}
\caption{\label{fig:gammap_star}
The contribution to the correlator $\gamma_p$ from local charge conservation superimposed onto elliptic flow from the model is compared to measurements from the STAR Collaboration \cite{Abelev:2009ac}. The dashed green line shows contributions from correlations from the hydrodynamic stage, while the dotted line represents correlations born in the cascade. The sum (solid blue line) is $\approx 10-15\%$ higher than the data. Thus, the combination of charge conservation and flow more than accounts for the observed correlation, which has been proposed as a signal of the chiral magnetic effect.
}
\end{figure}
Figure \ref{fig:gammap_star} compares model calculations of $\gamma_p$ to those from STAR \cite{Abelev:2009ac,Adamczyk:2013hsi}. The contribution from correlations from the cascade stage provide $\approx 60\%$ of $\gamma_p$ even though the represent only $\approx 40\%$ of the strength of the balance functions in Fig. \ref{fig:bf_allcents}. The larger role of the correlations from the cascade comes from their being more narrow, hence $\cos\Delta\phi$ is larger. The net correlation from the model calculations are 10-15\% larger than the STAR data over the range of centralities.

Given that the charge balance functions in Fig. \ref{fig:bf_allcents} for the centrality range of 40-50\% lie above the data, motivating the adjustment factor of 0.94 in Fig. \ref{fig:bf_phi1phi_star}, one would expect the model prediction of $\gamma_p$ to be high by approximately 6\%, since the normalization discrepancy would be due to more balancing charges lying outside the acceptance in the experiment than in the model, and only those correlations within the acceptance contribute to $\gamma_p$. Over-stating the flow would also lead to over-predictions of $\gamma_p$, but if that were the case, one would expect the reaction-plane-dependent balance functions of Fig. \ref{fig:bf_phi1phi_star} to have a discrepancy with the data. Another possibility would be for the multiplicity of the model to under-predict the true experimental situation. This was checked, and it seems unlikely this could be a 10\% discrepancy. Thus, after accounting for the difference in normalizations between the model and data for less central events, this analysis suggests that flow plus local charge conservation would predict values close to the upper limits of the experimental error bars. Given that the error bars include systematic error, it is not out of the question that practically all of the observed correlation, $\gamma_p$, derives from charge balance and flow. If the portion of the signal from the chiral magnetic effect were 10\% of the signal, one would need to explain away an even more significant over-prediction of the model. Additionally, the CME in isolation gives a negative balance function for out-of-plane pairs, whereas charge balance and elliptic flow lead to positive correlations, but with stronger positive correlations in-plane. The latter is what is observed, but this does not preclude the possibility that some of the difference derives from the CME. It does seem unlikely that the CME contribution could be larger than 10\% of the signal given the model-data comparisons in Figs. \ref{fig:bf_phi1phi_star} and \ref{fig:gammap_star}. Similar conclusions were generated by comparing to simpler parametric models of flow and charge conservation \cite{Schlichting:2010qia}, or to a very simple pion cascade model \cite{Pratt:2010zn}.

\section{Summary and Conclusions}
\label{sec:summary}

By incorporating a hadronic cascade into the hydrodynamics-only treatment from \cite{Pratt:2017oyf}, the treatment of correlations related to charge conservation is now based on a beginning-to-end state-of-the-art transport picture. The previous calculations of \cite{Pratt:2017oyf} ignored the hadron cascade, which was less realistic, though significantly less expensive numerically. The numerical expense of this approach was mainly due to the fact that correlations from the cascade portion mixed amongst colliding particles, which was handled by treating the output in the same manner as what is done in experiment. The combinatoric noise required calculations of the equivalent of $\approx 10$ million cascade events. This represented a few months of CPU time, which is not a particularly daunting cost, but does make it challenging to explore a high dimension parameter space.

The approach of combining a hydrodynamic model and cascade has become recognized as a ``best-practice'' approach to describing relativistic heavy-ion collisions. Given that the equation of state and charge susceptibilities are well determined by lattice gauge theory, there are few aspects of the model that one might alter that would significantly change the outcome. Initial collective flow \cite{Gelis:2013rba,Dusling:2010rm,Vredevoogd:2008id}, i.e. flow from before hydrodynamics is instantiated, is neglected here but should only affect the collective flow at the 5\% level for central or mid-central collisions. The viscosity in the hydrodynamic stage was set so that $\eta/s=1/4\pi$, which might be on the low end, but again, it is not clear that doubling the viscosity would significantly change the charge balance functions. More sophisticated hydrodynamic treatments include event-by-event fluctuations, or lumpy initial conditions. Although including or adjusting these various features would significantly change certain relevant observables, it is not expected that they would change charge balance functions by more than a few percent. A more important parametric choice here was for the diffusion constant for light quarks, which was taken from lattice gauge theory. However, the lattice values, which are functions of temperature, are somewhat untrustworthy due to the fact that their extraction from lattice requires an analytic continuation. The most uncertain parameter, and the one that most strongly affects the results is the choice of $\sigma_0$. This parameter represents the random distance, in spatial rapidity, that each charge has moved relative to its balancing charge at the initial time where hydrodynamics is invoked. For example, if the initial charges came from the decay of longitudinal flux tubes, one would expect the two charges to pull apart during the tunneling process that provide the energy for particle production. This parameter is known from \cite{Pratt:2017oyf} to significantly affect the kaon and proton balance functions binned by relative rapidity.

Here, results were shown for both identified and non-identified (aside from charge) particles, binned by relative rapidity and relative azimuthal angle. Results were compared to measurements from the STAR Collaboration for $\sqrt{s}_{NN}=200$ GeV/$c$ Au+Au collisions at RHIC. The model remarkably well described charge balance functions for unidentified particles binned by either relative azimuthal angle or relative pseudo-rapidity for a range of centralities. The only discrepancy seemed to be that for less central collisions, the experimental balance functions were modestly wider than those from the model. The model also well described balance functions for identified particles. The $p\bar{p}$, $K^+K^-$ and $\pi^+\pi^-$ balance functions were well described, aside from the $p\bar{p}$ balance function missing a dip at small relative momentum due to annihilation in the breakup stage. In fact, the shape of the $\pi^+\pi^-$ balance function binned by relative rapidity was marginally better reproduced than in the less sophisticated model of \cite{Pratt:2017oyf}. The one noticeable failure was in reproducing the $pK^-$ balance function, a shortcoming also seen in \cite{Pratt:2017oyf}. 

The fact that the $K^+K^-$ and $p\bar{p}$ balance functions are broader than the $\pi^+\pi^-$ balance functions, both in the data and in the model, suggests that a chemically equilibrated QGP was produced early in the collision. Because the strangeness and baryon susceptibilities, relative to the entropy density, stay nearly constant, one expects little contribution, or perhaps a negative contribution to the $K^+K^-$ and $p\bar{p}$ balance functions from late-stage production. Hence they are driven by the correlations that were generated in the early stage and significantly spread out in relative rapidity. In contrast, the $\pi^+\pi^-$ balance function is driven by the electric charge susceptibility which has a strong surge in the hadronization stage, and strong contributions from decay. These correlations tend to be much shorter range in coordinate space, which translates to narrower balance functions in relative rapidity. Indeed, these features were seen in the data, and were quantitatively reproduced by the model.  Although the failure of the model to describe the $pK^-$ balance function dampens the enthusiasm for claiming success, the preceding discussion suggests that relatively minor improvements might be sufficient to bring model results into agreement with experimental data.  We defer these improvements to a future study.

Another, more differential, set of charge balance functions involved constraining the first particle's azimuthal angle relative to the reaction plane. By plotting balance functions for unidentified particles binned by relative azimuthal angle subject to this constraint, one is given a highly detailed test of elliptic flow, and the dynamics of correlations stemming from local charge conservation. The reproduction of the experimental measurements was rather stunning. This success also translated into the moment $\gamma_p$, which had been suggested as a signal of the chiral magnetic effect (CME), and can be uniquely stated in terms of integrals of the reaction-plane-dependent balance functions. The model prediction of $\gamma_p$ over-predicted the experimental measurement by 10-15\%. Some of this over-prediction was expected given that the experimental balance functions seemed to spread more outside the acceptance, but most of the over-shoot remained unexplained, though the size of the discrepancy was not far outside the systematic error bars of the experiment. This result makes it difficult to imagine a situation where the CME contributions could be sufficiently substantial to be separated from the effects of local charge conservation super-imposed onto elliptic flow. 

Going forward, the main facet of the model that requires attention is baryon annihilation. This should give an extra dip at low relative momentum to the $p\bar{p}$ balance function. As for future analyses, it would be most interesting to study how well one can extract the diffusion coefficient from these models. Because of the unknown separation in relative spatial rapidity when hydrodynamics is initialized, $\sigma_0$, the widths of balance functions in rapidity is probably not a robust means to study the diffusion coefficient. However, the width in relative azimuthal angle seems promising, and will be the subject of a future study.

\appendix
\section{Modified Cooper Frye Formula}
\label{sec:CFappendix}

Rewriting Eq. (\ref{eq:dNfromdQ}) which describes the differential number of hadrons of species $h$ from a differential charge $dQ_a$,
\begin{eqnarray}
\label{eq:dNfromdQapp}
dN_h&=&n_hq_{h,a}(\chi^{-1})_{ab}dQ_b,
\end{eqnarray}
one can probabilistically choose whether or not to create a hadron of species $h$ with probability $dN_h$ for each sample charge that passes through the hyper surface. Using the Cooper-Frye formula, 
\begin{eqnarray}
\label{eq:CFapp}
dN_h&=&\frac{d^3p}{E_p}~p\cdot d\Omega~f_h(\vec{p}),
\end{eqnarray}
the momentum of said hadron was chosen by assuming the additional hadron's momentum was proportional to the right-hand side of Eq. (\ref{eq:CFapp}). This was performed by first transforming to the rest frame of the fluid element. In that frame 
\begin{equation}
\label{eq:CFrestframe}
dN_h=d\Omega_0 ~d^3p~f_h(\vec{p})\left[1+\vec{v_p}\cdot d\vec{\Omega}/d\Omega_0\right].
\end{equation}
Here, $\vec{v}_p$ is the velocity of a particle in this frame. If the hyper element is time-like, $|d\vec{\Omega}|<d\Omega_0$, the right-hand side of Eq. (\ref{eq:CFrestframe}) is positive for all $\vec{p}$. If $f_h$ is thermal, i.e. $f_h=e^{-E_p/T}$, a momentum can readily be chosen proportional to $d^3p~f_h$. Even if there are viscous corrections, one can adjust the generation of the momentum consistent with the deviations of the stress-energy tensor. For this study we follow the method described in \cite{Pratt:2010jt}. Next, to account for the weight,
\begin{equation}
\label{eq:weightapp}
w=1+\vec{v_p}\cdot d\vec{\Omega}/d\Omega_0,
\end{equation}
one can reflect the momentum about the $d\vec{\Omega}$ plane with a probability,
\begin{eqnarray}
\label{eq:probrefl}
P_{\rm reflect}&=&\left\{\begin{array}{cc}
0,& \vec{v_p}\cdot d\Omega>0\\
|\vec{v_p}\cdot d\Omega|/d\Omega_0,& \vec{v_p}\cdot d\vec{\Omega}<0
\end{array}\right. .
\end{eqnarray}
On average, the procedure would perfectly represent $dN_h$ from the Cooper-Frye formula, and because the choice of whether to produce the hadron was from Eq. (\ref{eq:dNfromdQapp}), the procedure would be perfectly consistent. However, an issue arises when the reflection probability in Eq. (\ref{eq:probrefl}) is negative, which is the same as saying the $dN_h$ would be negative according to Eq. (\ref{eq:CFrestframe}). For this treatment, the reflection probability was simply chose to be unity in such cases. This approximation could be overcome by consistently considering the case where cascade particles re-enter the hydrodynamic region by crossing the same hyper element. If such crossings were consistent with the phase space density expressed above, as would be the case if the hyper-surface were indeed chosen at a point where the phase space density in the cascade maintained a continuous phase space density, a simple procedure would be for such particles to reflect about the $d\vec{\Omega}$ plane. The removal of the incoming cascade particle would represent the negative contribution of the Cooper-Frye formula and the reflected particle would account for the part of the weight in Eq. (\ref{eq:weightapp}) that exceeds 2, i.e. when $\vec{v_p}\cdot d\vec{\Omega}/d\Omega_0>1$.

For this paper, the reflection of cascade particles was not performed. Because the reflection does not change the energy of the particle, in the fluid frame, and because the reflection does not create or destroy charges, the approximation does not violate charge conservation, or energy conservation. However, it does represent a small violation of momentum conservation, and in a different frame this would translate into a violation of energy conservation. Fortunately, such reflections affect less than one percent of the particles in high-energy collisions. This is because most particles are emitted through time-like hyper-surface elements, and even for those space-like hyper-surface elements that one encounters, only a small fraction of the momentum space has a negative contribution in the Cooper-Frye formula. Finally, because this study is concerned with charge conservation, this choice of approximation should be especially warranted. 

Because a hadron produced through the procedure thus represents the situation in the fluid frame, it is then boosted to the laboratory frame to complete the procedure.

\begin{acknowledgments}
This work was supported by the Department of Energy Office of Science through grant number DE-FG02-03ER41259 and through grant number DE-FG02-87ER40328. 
\end{acknowledgments}



\begin{thebibliography}{99}
\bibitem{WestfallAcceptance}
 Routines for modeling the efficiency of the STAR detector were generously provided by Gary Westfall.
\end{thebibliography}
\end{document}